\documentclass[%
reprint,                                                                       
prb,                                                                           
]{revtex4-1}
\usepackage{mystyle}
\newcommand{\etsf}{European Theoretical Spectroscopy Facilities (ETSF)}
\newcommand{\qub}{Atomistic Simulation Centre, School of Mathematics and Physics, 
                  Queen's University Belfast, 
                  Belfast BT7 1NN, UK}

\begin{document}

\title{
    Projected Equations of Motion Approach to Hybrid Quantum/Classical
    Dynamics in Dielectric-Metal Composites
}

\author{Ryan J. McMillan}
\email[]{rmcmillan05@qub.ac.uk}
\affiliation{\qub}
\affiliation{\etsf}
\author{Lorenzo Stella}
\affiliation{\qub}
\affiliation{\etsf}
\author{Myrta Gr\"uning}
\affiliation{\qub}
\affiliation{\etsf}

\date{\today}

\begin{abstract}
    We introduce a hybrid method for dielectric-metal composites that describes
    the dynamics of the metallic system classically whilst retaining a quantum
    description of the dielectric. The time-dependent dipole moment of the
    classical system is mimicked by the introduction of projected equations of
    motion (PEOM) and the coupling between the two systems is achieved through an
    effective dipole-dipole interaction. To benchmark this method, we model a
    test system (semiconducting quantum dot-metal nanoparticle hybrid). We
    begin by examining the energy absorption rate, showing agreement between
    the PEOM method and the analytical rotating wave approximation (RWA)
    solution. We then investigate population inversion and show that the
    PEOM method provides an accurate model for the interaction under
    ultrashort pulse excitation where the traditional RWA breaks down.
\end{abstract}
\maketitle

\section{Introduction\label{sec:intro}}

The electronic structure and quantum dynamics of a system
can be modelled using several approaches based on, e.g., wave function
methods~\cite{landau1977quantum}, Green's
functions~\cite{stefanucci2013nonequilibrium,nolting2009fundamentals,marini_yambo:_2009}, density
matrix theory~\cite{rand_lectures_2010,boyd_nonlinear_2008} or density
functional theory (DFT)~\cite{martin2004electronic,marques2012fundamentals}. 
In practice, to model larger and larger electronic systems,  
high-performance computing facilities along with optimized algorithms are continually
developed. To improve the scaling of the algorithm, hybrid approaches have been devised 
to break down the computational complexity of composite systems which 
include a small subsystem --- still amenable of
a fully quantum-mechanical treatment --- and a larger environment --- which 
is dealt with a lower level of approximation, most often classical.
Examples of such composites include solvated
molecules~\cite{cossi_ab_1996,cossi_new_2002,tomasi_quantum_2005},
protein-ligand
interactions~\cite{murphy_mixed_2000,hensen_combined_2004,grater_protein/ligand_2005}
and semiconductor-metal nanoparticle
hybrids~\cite{lin_plasmonic_2013,singh_enhancement_2013,zhang_semiconductor-metal_2006,artuso_strongly_2010,artuso_hybrid_2012,paspalakis_control_2013,kosionis_optical_2013,cheng_coherent_2007,li_optical_2012,singh_enhancement_2013}.  In all these
cases, we are more interested in the dynamics of the smaller subsystem and we
look at the environment as a source of unavoidable perturbations.

Such hybrid approaches rely on the possibility to
separate the composite system into two or more components whose dynamics are
solved using different levels of approximation and to treat the residual 
interaction between the subsystems in an appropriate way. For example, a continuum
solvation model (such as the polarizable continuum model) may be used in the
solvated molecule problem where the molecule is treated using quantum mechanics
(QM) 
and the solvent treated as a dielectric
continuum, the interaction being electrostatic in
nature~\cite{cossi_ab_1996,cossi_new_2002,tomasi_quantum_2005,corni_equation_2015}. Various
quantum mechanics/molecular mechanics (QM/MM) approaches have also been applied to
model the protein-ligand interaction. In these cases, the ligand is treated using
QM while the protein environment via MM and
the potentials associated with the protein's molecular
make-up is approximated by means of classical force fields~\cite{murphy_mixed_2000,hensen_combined_2004,grater_protein/ligand_2005}.

Hybrid methods have also been applied to model the coupling between molecules and metal
nanoparticles (MNPs) upon optical excitation. For small MNPs, the composite system
can still be treated fully quantum mechanically~\cite{zuloaga_quantum_2009}. For
larger MNPs, classical electrodynamics is employed to model the MNP dynamics
whereas a quantum description of the molecule is retained. In this case,
the interaction between the MNP and the molecule is modeled through an
effective electromagnetic coupling. These hybrid approaches make use of numerical methods such
as the finite-difference time domain (FDTD)
to solve the classical electrodynamics problem --- namely,
the Maxwell's equations --- while the dynamics of the 
molecular electrons are solved by means of time-dependent
DFT. The overall dynamics are made self-consistent
by including the electromagnetic field generated by the MNP into the molecular evolution
and \emph{vice versa}~\cite{chen_classical_2010,coomar_near-field:_2011,gao_communication:_2013,sakko_dynamical_2014}. 

In this work we propose an alternative, simpler and much less computationally expensive method 
that avoids the solution of Maxwell's equations when the near--field effects in the electromagnetic 
coupling between the MNP and the quantum system (e.g., a molecule or a quantum dot) are negligible.
To this end, we shall present a generalized model for treating the time-dependent interaction
between a quantum system (QS) and classical system (CS) coupled through an
electromagnetic field. The interaction is considered in the dipole-dipole
approximation
within the quasi-static limit. The dynamics of the QS are described via the
density matrix master equation involving an effective field which depends on the
time-dependent dipole moment of the CS. We note here that whilst we employ density
matrix theory for the quantum dynamics, the method is general and can be
applied to any time-dependent quantum mechanical approach such as those
mentioned in the opening paragraph.
The CS is modeled using classical electrodynamics in the linear response
regime where the time-dependent dipole moment is reproduced by the
introduction of a set of auxiliary degrees of freedom.
These degrees of freedom enter into a set of projected equations 
of motion (PEOM) and are constrained by modeling the frequency-dependent
polarizability of the CS.

As a testbed, we consider the hybrid system consisting of a semiconducting
quantum dot (SQD) and MNP. In particular, the SQD is treated as an abstract
two-level QS while the MNP is modelled as a gold nanosphere. This system has been
studied
extensively~\cite{zhang_semiconductor-metal_2006,artuso_strongly_2010,artuso_hybrid_2012,paspalakis_control_2013,kosionis_optical_2013,cheng_coherent_2007,li_optical_2012,singh_enhancement_2013}
because it can be solved analytically by means of the rotating wave approximation
(RWA). For continuous wave excitation, we show that this analytical benchmark for the 
energy absorption is correctly retrieved by the proposed hybrid approach.
Pulsed excitations are also examined and agreement between the proposed method and the 
RWA approximation is shown if picosecond pulses are used. However, for a femtosecond pulse
the RWA breaks down and an approach like the proposed method must be preferred.

The paper is organised as follows: In Section~\ref{sec:methods}, we
describe the dipole-dipole interaction between the QS and CS and derive the
PEOM method for treating the time-dependent
dipole moment of the CS. The method is applied to a simple SQD-MNP system in
Section~\ref{sec:testbed1} and the results for energy absorption rates and
population inversion are compared with those from semi-analytical
approximations. Finally, the conclusions are presented in
Section~\ref{sec:conclusions}.

\section{\label{sec:methods}The PEOM Method}

\begin{figure}[ht]
    \centering
    \includegraphics[width=0.8\columnwidth]{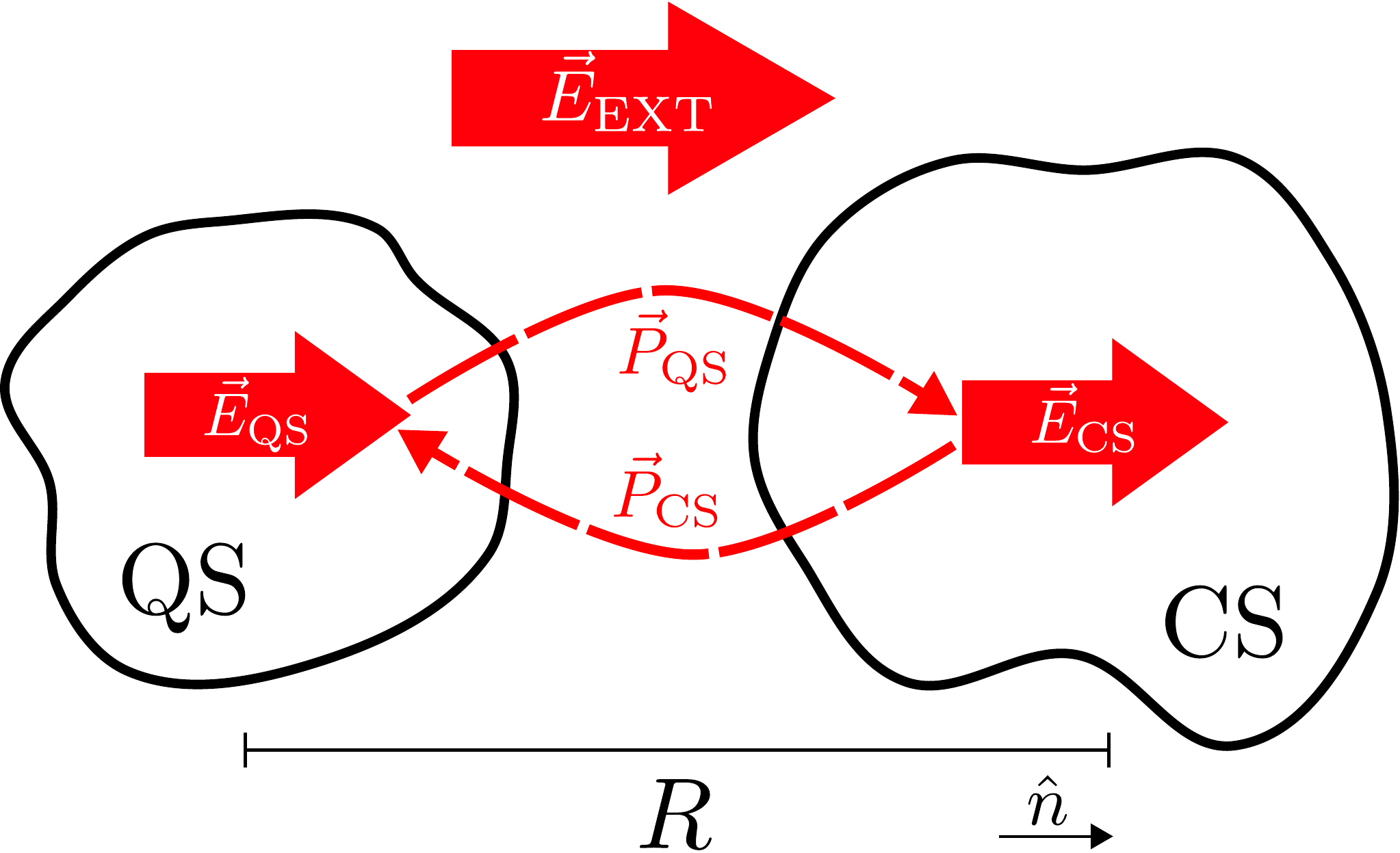}
    \caption{Schematic diagram showing the dipole-dipole
        interaction between a QS and a CS,
        separated by a distance $R$.
        When an external field, $\veext$, is applied, a dipole, $\vpqs$, is
        induced in the QS thus generating a field. The CS thus experiences this
        dipole field in addition to the external field, and we denote the total
        field felt by the CS as $\vecs$. Similarly, due to $\vecs$, a dipole field is generated
        in the CS which is in turn felt by the QS in addition to $\veext$, and we
        denote the total field felt by the QS as $\veqs$. In this way, the QS
    and CS dynamics are coupled through the external field.}
    \label{fig:two_particles}
\end{figure}

We consider a QS and a CS
separated by a distance, $R$. An external field, $\veext(t)$, is applied inducing
a dipole-dipole interaction
between the two systems (see \figref{fig:two_particles}). To simplify the
notation, we assume that the QS and CS are isotropic media, though the method
can be easily generalized to the anisotropic case. We write
$\veext(t)\equiv\eext(t)\ve$ and denote the unit vector pointing along the
line separating the centers of the particles as $\hat n$. The fields felt by
the QS and CS are then,
respectively~\cite{jackson_classical_1962,schmitt_preparation_1999,zhang_semiconductor-metal_2006},
\begin{subequations}
    \noeqref{eq:eqs,eq:ecs}
    \begin{align}
        \veqs(t) &= \eext(t)\ve + \frac{\pcs(t)}{\epsb R^3}\g \ , \label{eq:eqs}\\
        \vecs(t) &= \eext(t)\ve + \frac{\pqs(t)}{\epsb R^3}\g \ , \label{eq:ecs}
    \end{align}
    \label{eq:fields1}
\end{subequations}
where $\pcs(t)$ ($\pqs(t)$) is the total dipole moment of
the CS (QS). $\epsb$ is the dielectric constant of the background medium and 
\begin{equation}
    \g=3\hat n\left( \ve \cdot \hat n \right) - \ve \ .
    \label{eq:g_def}
\end{equation}
The form of the fields here assumes only a dipole interaction. This is valid if
$R$ is sufficiently large, but can be generalised to take into account
higher multipole interactions as shown in Ref.~\onlinecite{Yan2008a}.

For demonstration, we shall presently use the density matrix approach to describe the
quantum system, although the method is applicable to any time-dependent
model. In the density matrix formalism, the dipole moment of the QS is given by
\begin{equation}
    \noeqref{eq:pqs}
    \vpqs(t) = \tr{\bm\rho(t)\vmu} \ ,
    \label{eq:pqs}
\end{equation}
where $\tr{\cdots}$ is the matrix trace operator, $\vmu$ is the dipole moment operator matrix and $\bm\rho$ is the QS
density matrix which evolves in time due to the field, $\veqs(t)$, via
the following master equation,
\begin{equation}
    \dot{\bm\rho} = -\frac{i}{\hbar} \left[\bm H_0, \bm \rho\right] +
    \frac{i}{\hbar} \veqs(t)\cdot \left[\vmu,\bm\rho\right] +
    \eta\left(\bm\rho\right) \ .
    \label{eq:rho_eom}
\end{equation}
In Eq.~\eqref{eq:rho_eom}, $\bm H_0$ is the Hamiltonian of the unperturbed QS
and the interaction energy with the time-dependent field $\veqs(t)$ is treated within the electric
dipole approximation (second term). $\eta\left(\bm\rho\right)$ is an
additional function which can be used to model phenomenological effects not
included in the Hamiltonian such as non-radiative decay (see Ref.~\onlinecite{boyd_nonlinear_2008}, for example).

We assume that the CS has a frequency dependent polarizability $\alpha(\omega)$
which is known, e.g. by experiment or \textit{ab-initio} calculations.  Its
dipole moment can then be described (in the linear response regime)
via~\footnote{Gaussian units are assumed throughout the paper.}
\begin{equation}
    \vpcs(\omega) = \epsb  \alpha(\omega)\vecs(\omega) \ .
    \label{eq:pcs_omega}
\end{equation}
In the time domain, the dipole moment
is written in terms of the response function $\alpha(t)$,
\begin{equation}
    \vpcs(t) = \epsb \int_{-\infty}^{t}\alpha(t-t')\vecs(t')dt' \ .
    \label{eq:P_2}
\end{equation}
Using a coupled iterative technique, we could then solve Eq.~\eqref{eq:rho_eom}
numerically to obtain the time-dependent response of the QS to
the effective field $\veqs(t)$. 
However, computing the integral in Eq.~\eqref{eq:P_2} 
at each time-step in the solution is cumbersome and the values of
$\vecs(t)$ and $\alpha(t)$ for each time-step must be held in memory which may not be
feasible for long simulations.
This leads to the main component of the PEOM method, an alternative to
calculating Eq.~\eqref{eq:P_2} directly by following a time-convolutionless
scheme inspired by Ref.~\onlinecite{stella_generalized_2014}.

We introduce $N$ complex auxiliary degrees of freedom, $\{s_k(t)\}$ for
$k=1,2,\ldots,N$, which satisfy the following EOMs
\begin{equation}\label{eq:s_EOM}
    \dot{\vs}_k = -\left( \gamma_k + \imagn\omega_k \right)\vs_k + \imagn\epsb
    \vecs(t)\
    ,
\end{equation}
and assume that $\vpcs(t)$ can be written as
\begin{equation}\label{eq:P_2_approx}
    \vpcs(t) = \sum_{k=1}^{N} c_k \real{\vs_k(t)} \ ,
\end{equation}
so that the memory-dependent integral in Eq.~\eqref{eq:P_2} is replaced with an
expansion over the functions $\vs_k$ found by solving the differential
equations in Eq.~\eqref{eq:s_EOM}. As these differential equations no longer
contain a time-convolution (i.e., they are ``memoryless''), they can be
efficiently integrated by using standard iterative algorithms, e.g., the
Runge-Kutta fourth-order method. All that
is required is to find suitable values for the (real) parameters $\{c_k,
\gamma_k, \omega_k\}$ in Eq.~\eqref{eq:s_EOM}.
The formal solution of Eq.~\eqref{eq:s_EOM} is
\begin{equation}
    \vs_k(t) = \epsb \int_{-\infty}^{t}\imagn\ e^{-\left( \gamma_k + \imagn\omega_k
    \right)\left( t-t' \right)} \vecs (t') dt' \ .
    \label{eq:s_k}
\end{equation}
Substituting the real part of Eq.~\eqref{eq:s_k} into Eq.~\eqref{eq:P_2_approx} and
rearranging yields
\begin{align}
    \vpcs(t) = \epsb
    &\int_{-\infty}^{t} 
    \left( 
        \sum_{k=1}^{N} c_k e^{-\gamma_k (t-t')}\sin[\omega_k(t-t')]
    \right) \notag \\
    & \times
    \left( \vecs(t') \right) dt' \ ,
\end{align}
and comparing with Eq.~\eqref{eq:P_2} we see that
\begin{equation}
    \alpha(t) = 
        \sum_{k=1}^{N} c_k e^{-\gamma_k t}\sin(\omega_k t) \ .
    \label{eq:chi_t}
\end{equation}
Then taking the Fourier transform of
Eq.~\eqref{eq:chi_t} (using the causality condition) gives
\begin{equation}
    \alpha(\omega) = \sum_{k=1}^{N}\frac{c_k}{2}
    \left[ \frac{1}{\omega+\omega_k +\imagn\gamma_k}
    -\frac{1}{ \omega-\omega_k +\imagn\gamma_k}\right] \ .
    \label{eq:chi_approx}
\end{equation}
Hence, the parameters $\{c_k,\gamma_k,\omega_k\}$ may be found by fitting
the frequency-dependent polarizability of the CS (which is known)
to the fitting functions on the RHS of Eq.~\eqref{eq:chi_approx}
(e.g. using the least squares method as done in this work).

\section{The Quantum Dot-Metal Nanoparticle System\label{sec:testbed1}}

To test the PEOM method proposed in
\secref{sec:methods}, we use a semiconducting
quantum dot (SQD) as the QS and a metal nanoparticle (MNP) as the
CS since this hybrid system has been widely
studied~\cite{zhang_semiconductor-metal_2006,artuso_strongly_2010,artuso_hybrid_2012,paspalakis_control_2013,kosionis_optical_2013,cheng_coherent_2007,li_optical_2012,singh_enhancement_2013}
and some properties can be obtained analytically.
In particular, we look at the energy absorption rate (EAR) and
population inversion, which are associated with continuous and pulsed wave
excitation respectively. 

From Eq.~\eqref{eq:fields1}, when an external field $\eext$ is applied, the
fields felt by the SQD and MNP respectively are
\begin{align}
    \esqd &= \eext + g\frac{\pmnp}{\epsb R^3} \ , \label{eq:esqd}\\
    \emnp &= \eext + g\frac{\psqd}{\epsb R^3} \ .
\end{align}
We have taken the external field to be polarized along the line connecting the
centers of the particles, allowing us to drop the vector notation and set $g=2$
(see Eq.~\eqref{eq:g_def}). 

The SQD is
treated as a 2-level atomic system giving rise to a $2\times 2$ density matrix
with elements that can be written
as~\cite{boyd_nonlinear_2008,zhang_semiconductor-metal_2006}
\begin{equation}
    \begin{cases}
        \dot{\Delta} &= -\frac{4\mut}{\hbar}\esqd(t)\imag{\rho_{21}} -
        \Gamma_{11}(\Delta - 1)   \\
        \dot\rho_{21} &= -\left( \Gamma_{21}+\imagn\omega_0
        \right)\rho_{21} + \imagn\frac{\mut}{\hbar}\esqd(t) \Delta
        \label{eq:eoms}
    \end{cases} \ ,
\end{equation}
where $\Delta(t)=\rho_{11}(t)-\rho_{22}(t)$ is the population difference between the
ground and excited states with frequency difference $\omega_0$ which is known
as the exciton frequency. $\Gamma_{11}$ and $\Gamma_{21}$ are the population
decay and dephasing rates of the system respectively. The SQD is assumed to be
a
dielectric sphere with dielectric constant $\epss$ and so it has a screened dipole matrix element
$\mut=\mu_{21}/\epseff$ where $\mu_{21}$ is the bare dipole matrix element
and $\epseff=\frac{2\epsb+\epss}{3\epsb}$.~\cite{batygin_problems_1978}

For comparison with previous literature, the MNP is taken to be a gold sphere
of radius $a$ and its polarizability is approximated by the Clausius-Mossotti
formula,
\begin{equation}
    \alphamnp(\omega) = a^3 \frac{\epsm(\omega)-\epsb}{
    \epsm(\omega) + 2\epsb} \ ,
    \label{eq:chi_sphere}
\end{equation}
where $\epsm(\omega)$ is the frequency-dependent dielectric function of the
bulk metal~\cite{landau_electrodynamics_1984} (we use the analytical model for
bulk gold as given by Etchegoin et al.~\cite{etchegoin_analytic_2006}). Note
that there are no fundamental reasons for using an analytical expression for
the polarizability. For example, $\alphamnp(\omega)$ may instead be extracted from
experimental data or computed using a first-principles approach.

We take the SQD system parameters from
Ref.~\onlinecite{zhang_semiconductor-metal_2006}.
The dielectric
constant is taken to be $\epss=6$ with transition dipole moment $\mu=0.65e$~nm and
exciton energy $\hbar\omega_0=2.5$~eV close to the plasmon peak of the gold MNP.
The decay and dephasing times are given by
$\Gamma_{11}^{-1}=0.8$~ns and $\Gamma_{21}^{-1}=0.3$~ns. We assume the
background medium is a vacuum so that $\epsb=1$.

\subsection{Energy Absorption Rate}

We first look at the EAR of the hybrid system which is a steady-state property,
found by considering the response to the following field,
\begin{equation}
    \eext(t) = E_0 \cos(\omega_L t) .
    \label{eq:plane_wave}
\end{equation}
In this case, Eq.~\eqref{eq:eoms} can be solved analytically within the RWA as
shown in, e.g.,
Refs~\onlinecite{zhang_semiconductor-metal_2006,artuso_strongly_2010,artuso_hybrid_2012}.
In the RWA, the off-diagonal density matrix elements are first separated into slowly and quickly
oscillating components,
\begin{subequations}
    \noeqref{eq:rho21,eq:rho12}
    \begin{align}
        \rho_{21}(t) &= \bar\rho_{21}(t)e^{-\imagn\omega_L t} \ , \label{eq:rho21} \\
        \rho_{12}(t) &= \bar\rho_{12}(t)e^{\imagn\omega_L t} \ , \label{eq:rho12}
    \end{align}
    \label{eq:rho_bar}
\end{subequations}
where $\bar\rho_{21}(t)$ and $\bar\rho_{12}(t)$ are assumed to vary on a much larger
timescale than $2\pi/\omega_L$. The RWA assumes
that $\omega_L\approx\omega_0$, neglecting terms oscillating at frequencies far
from $\omega_0$, so that the following modified EOMS can be obtained
from Eq.~\eqref{eq:eoms},
\begin{equation}
    \begin{cases}
        \dot\Delta &= 4\imag{\left( \frac{\omegaeff}{2}+G\bar\rho_{21} \right)\bar\rho_{12}}
        + \Gamma_{11}\left( 1-\Delta \right) \\
        \dot{\bar\rho}_{21} &= \left[ \imagn\left( \omega_L-\omega_0 +G\Delta
        \right)-\Gamma_{21} \right]\bar\rho_{21} + \imagn\frac{\omegaeff}{2}\Delta 
    \label{eq:2ls_eoms}
    \end{cases}
    \ ,
\end{equation}
where
\begin{subequations}
    \noeqref{eq:omegaeff1,eq:G}
    \begin{align}
        \omegaeff &=  \Omega_0 \left[ 1 +
        \frac{g}{R^3}\alphamnp(\omega_L) \right] \ ,\label{eq:omegaeff1}\\
        G &= \frac{g^2\mut^2 }{\hbar\epsb R^6}\alphamnp(\omega_L) \ , \label{eq:G}
    \end{align}
    \label{eq:omegaeff}
\end{subequations}
with $\Omega_0 = \mut E_0/\hbar$ being the Rabi frequency of the
isolated SQD.

The EAR of the
SQD is defined as~\cite{artuso_strongly_2010}
\begin{equation}
    \qsqd = \frac{1}{2}\hbar\omega_0\Gamma_{11}\left( 1-\Deltass \right)\ ,
    \label{eq:qsqd}
\end{equation}
where $\Deltass$ is the value of $\Delta(t)$ when a steady-state has been
reached, while the EAR of the MNP is~\cite{artuso_strongly_2010}
\begin{equation}
    \qmnp = \left\langle\int j\cdot \emnpin\ dV \right\rangle  \ ,
    \label{eq:qmnp}
\end{equation}
where $j = \frac{\partial}{\partial t} (\pmnp(t)/V)$ is the current density in
the MNP which has volume $V$ and $\emnpin$ is the field inside the MNP. Within
the RWA, it can be shown that $\qmnp$ depends on $\rhoss$, the steady state
value of $\bar\rho_{21}(t)$~\cite{artuso_strongly_2010}. The total EAR of the system is then
$Q=\qmnp+\qsqd$. An analytical solution for $\Deltass$ and
$\rhoss$ can be obtained by setting the L.H.S. of
Eq.~\eqref{eq:2ls_eoms} equal to zero (see
Ref.~\onlinecite{zhang_semiconductor-metal_2006} for example). 

As an alternative to the above RWA solution, we numerically solve the original
EOMs in Eq.~\eqref{eq:eoms} using the PEOM method where $\pmnp(t)$, which appears
in the expression for $\esqd(t)$, is approximated using
Eq.~\eqref{eq:s_EOM} and Eq.~\eqref{eq:P_2_approx}. The fitting parameters,
$\{c_k,\gamma_k,\omega_k\}$, are obtained from a least-squares fit of
$\alphamnp(\omega)$ to the model in Eq.~\eqref{eq:chi_approx} over the range 0-20~eV
which required $N=12$ fitting functions for sufficient accuracy.

\begin{figure}[ht]
    \centering
    \includegraphics[width=\columnwidth]{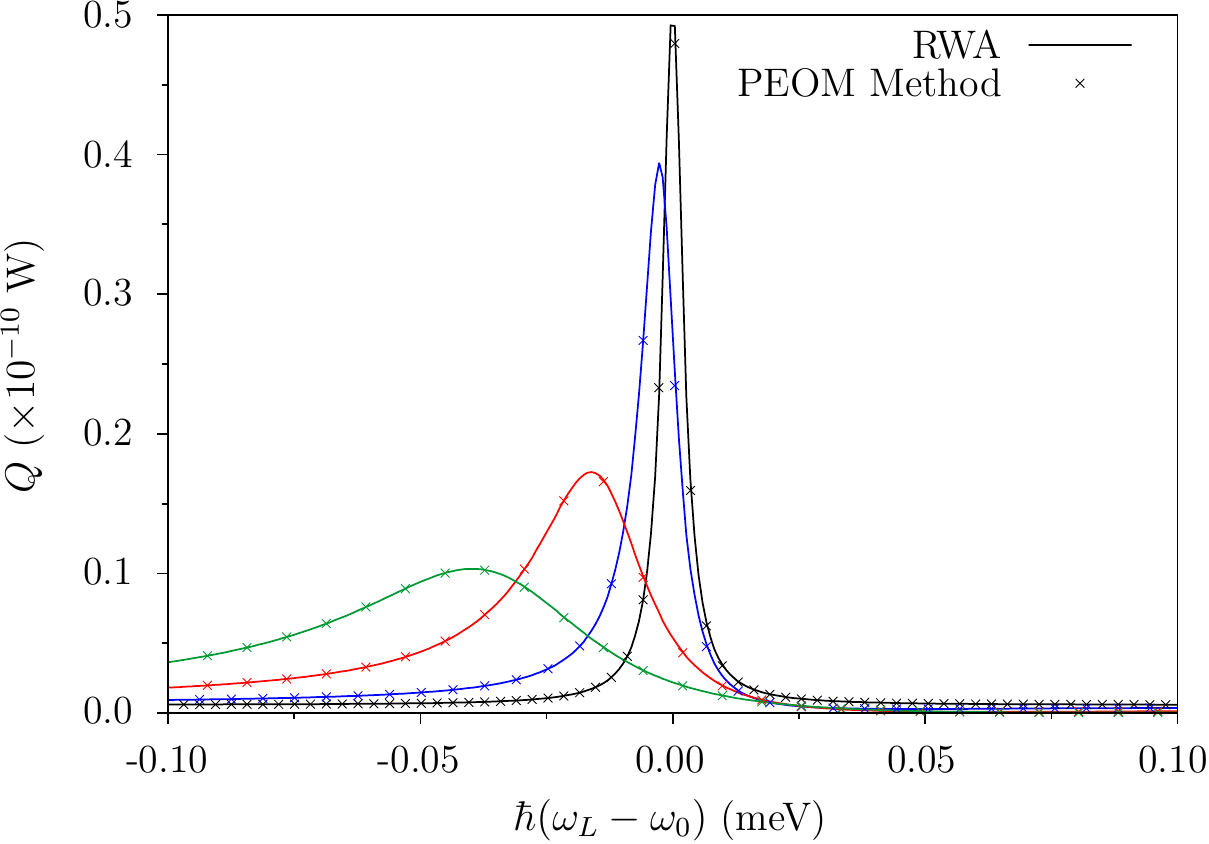}
    \caption{Energy absorption rate, $Q$, for a SQD-MNP system with
    separation distances $R=13$~nm (green), 15~nm (red), 20~nm (blue) and 80~nm
    (black).
    The solid
lines represent the steady-state analytical solution in the RWA while the crosses are
the results from the PEOM method.}
    \label{fig:Q_low_compare}
\end{figure}

\figref{fig:Q_low_compare} shows the total EAR for the hybrid system
as a function of the laser detuning (field intensity $I_0$=1~\wcm) for various separation distances of the two
particles. We can see the expected quenching of $Q$ and the red-shift of the
hybrid exciton energy as the particles are brought together as described in
Ref.~\onlinecite{zhang_semiconductor-metal_2006}. The analytical RWA
solutions are shown in
solid lines while the crosses are the results taken from the PEOM
method. In this case, we see perfect agreement between the two methods due
to the validity of the RWA for the case of a sinusoidal external field with
frequency very close to resonance with the SQD exciton frequency.
We now turn our attention to short-pulse excitation to
demonstrate a case where the RWA cannot be used.

\subsection{Population Inversion and Breakdown of the RWA}

Population inversion occurs when the SQD is
excited from the ground state to the excited state so that $\rho_{11}=0$,
$\rho_{22}=1$ and is associated with short laser pulses (see, for example,
Ref.~\onlinecite{stievater_rabi_2001}).
We consider a
pulsed external field given by
\begin{equation}
    \eext(t) = E_0 f(t) \cos(\omega_L t) \ ,
\end{equation}
where $f(t)$ is a dimensionless pulse envelope. The pulse area for an isolated
SQD is defined as
\begin{equation}
    \theta = \Omega_0\int_{-\infty}^{\infty} f(t) dt \ ,
    \label{eq:pulse_area}
\end{equation}
and it is known that population inversion occurs at the end
of the pulse for $\theta=(2n+1)\pi$
($n=0,1,2,\ldots$)~\cite{wang_decoherence_2005}.

We shall use a hyperbolic secant envelope defined by
\begin{equation}
    f(t) = \sech\left( \frac{t-\tau_0}{\tau_p} \right) \ ,
    \label{eq:sech}
\end{equation}
where $\tau_0$ is the center of the pulse and $\tau_p$ characterizes the pulse
width. We choose the central frequency, $\omega_L$, to be resonant with the
exciton frequency, i.e.
$\hbar\omega_L=\hbar\omega_0=2.5$~eV, and we describe the pulse shape in terms of the number of
cycles, $n$, by defining the pulse duration as
\begin{equation}
    T = \frac{4\pi}{\omega_L}n \ , 
    \label{eq:pulse_length}
\end{equation}
and choosing
\begin{equation}
    \tau_0 = T/2 \ , \quad \tau_p = T/30 \ .
    \label{eq:tau_0_tau_p}
\end{equation}
In this way we ensure the maximum amplitude, $E_0$, is achieved at the center
of the pulse and that the external field is sufficiently close to zero at $t=0$
and $t=T$ for the values of $E_0$ considered here.

For the sech pulse in Eq.~\eqref{eq:sech}, it can easily be shown from
Eq.~\eqref{eq:pulse_area} that for an isolated SQD, $\theta=\pi\Omega_0\tau_p$
and then a pulse of given duration can be described in terms of the pulse area by
choosing the following field amplitude,
\begin{equation}
    E_0 = \frac{\hbar\theta}{\pi\mut\tau_p} \ .
    \label{eq:theta_sech}
\end{equation}
In Ref.~\onlinecite{paspalakis_control_2013} it was shown that the pulse area
for an SQD when coupled to the MNP may be written approximately as $\theta =
\pi|\omegaeff|\tau_p$ so that Eq.~\eqref{eq:theta_sech} becomes
\begin{equation}
    E_0 = \hbar\theta \left( \pi\mut\tau_p \left|1 +
    \frac{g}{R^3}\alphamnp(\omega_L)\right| \right)^{-1} \ .
    \label{eq:theta_sech_R}
\end{equation}
In particular, it was stated that for short pulses ($\tau_p\sim0.1$~ps)
with amplitude given by Eq.~\eqref{eq:theta_sech_R}, the resulting dynamics
should be independent of $R$ as the influence of the parameter $G$
(see Eq.~\eqref{eq:G}) becomes weaker.

In previous studies relating to pulsed excitations in SQD-MNP systems, the
time-scales have generally been limited to relatively long pulses. For example in
Ref.~\onlinecite{sadeghi_coherent_2010,sadeghi_inhibition_2009}, the external
field is switched on over tens of nanoseconds while in
Ref.~\onlinecite{paspalakis_control_2013,anton_plasmonic_2012}, picosecond
pulses are used. In such cases, the population dynamics can be found by solving
the RWA EOMs in Eq.~\eqref{eq:2ls_eoms} but replacing $\omegaeff$ with the
time-dependent form,
\begin{equation}
    \Omega(t) = f(t) \omegaeff \ ,
    \label{eq:omegaeff_t}
\end{equation}
giving
\begin{equation}
    \begin{cases}
        \dot\Delta &= 4\imag{\left( \frac{\Omega(t)}{2}+G\bar\rho_{21} \right)\bar\rho_{12}}
        + \Gamma_{11}\left( 1-\Delta \right) \\
        \dot{\bar\rho}_{21} &= \left[ \imagn\left( \omega_L-\omega_0 +G\Delta
        \right)-\Gamma_{21} \right]\bar\rho_{21} + \imagn
        \frac{\Omega(t)}{2}\Delta 
    \label{eq:2ls_eoms_pulse}
    \end{cases}
    \ .
\end{equation}
The use of the RWA and slowly-varying envelope approximations respectively
imply that solutions to~\eqref{eq:2ls_eoms_pulse} are only valid if
${\bar\rho}_{21}(t)$ and $f(t)$ vary much more slowly than $2\pi/\omega_L$.
Recalling that $\hbar\omega_L=2.5$~eV, we therefore require the pulse duration
to be much greater than $\sim2$~fs. 

\begin{figure}[ht]
    \centering
    \begin{minipage}{0.9\columnwidth}
        \centering
        \includegraphics[width=1.0\columnwidth]{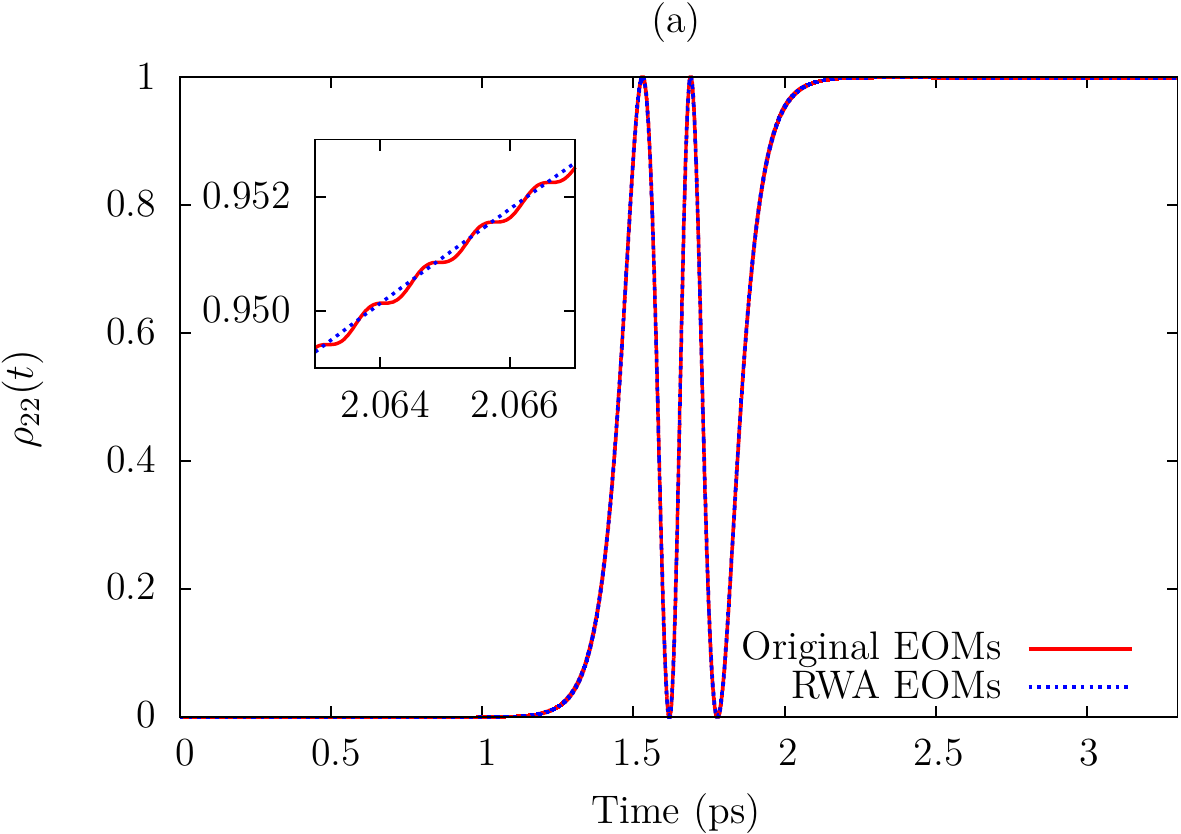}
        \vspace{0.2cm}
    \end{minipage}
    \begin{minipage}{0.9\columnwidth}
        \centering
        \includegraphics[width=1.0\columnwidth]{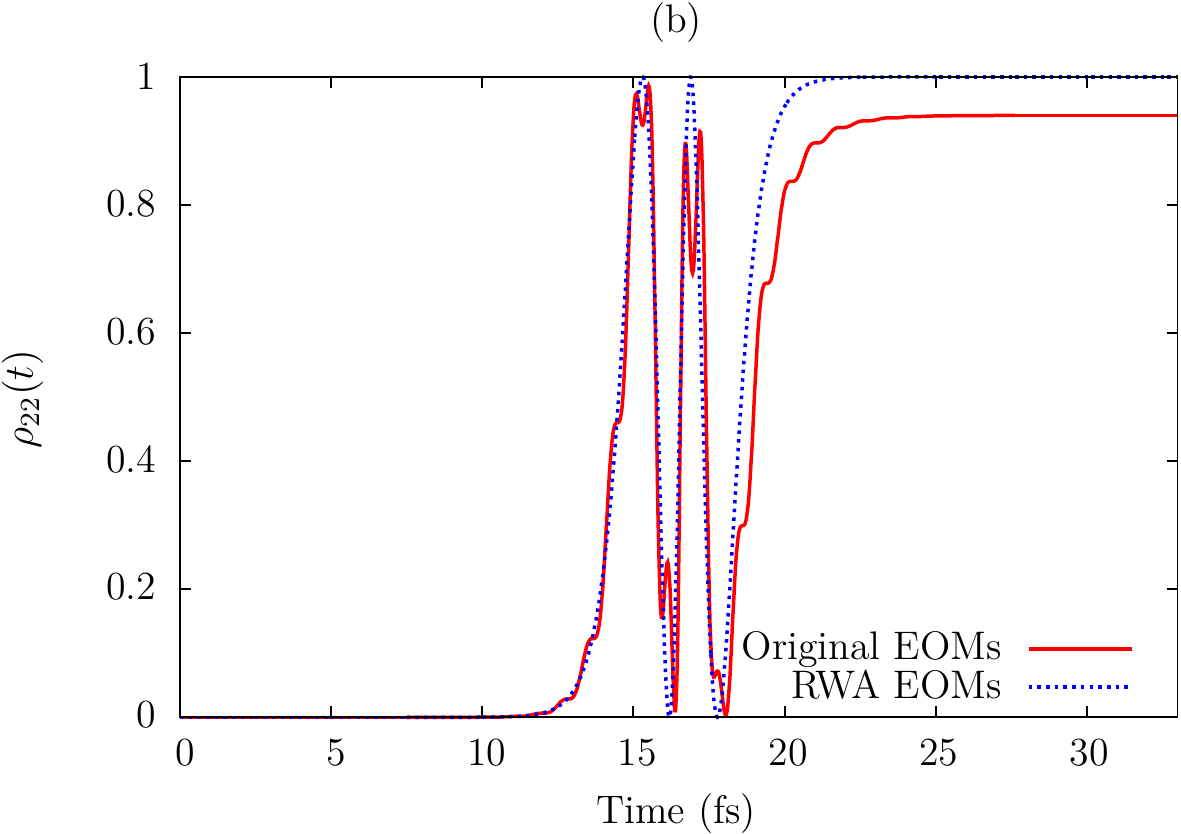}
    \end{minipage}
    \caption{Excited state population dynamics for an isolated SQD
            ($R\rightarrow\infty$) interacting with a sech pulse of area $5\pi$
            according to Eq.~\protect\eqref{eq:theta_sech}. The solid red line is the
            numerical solution to the original EOMs in Eq.~\protect\eqref{eq:eoms}
            while the dashed blue line is the solution to the modified RWA EOMs
            in Eq.~\protect\eqref{eq:2ls_eoms_pulse}. (a) Dynamics for a 1000-cycle
        pulse corresponding to $\tau_p\approx 0.11$~ps. (b) Dynamics for a
            10-cycle pulse corresponding to $\tau_p\approx 1.1$~fs.}
    \label{fig:rwa_compare}
\end{figure}

Indeed, it is known that the RWA is not reliable for
ultrashort (femto- and subfemto-second)
pulses.~\cite{meier_coherent_2007,PhysRevA.52.3082,yang_ultrafast_2015} This
is demonstrated in
Fig.~\ref{fig:rwa_compare} where
we compare the solution of the original EOMs in
Eq.~\eqref{eq:eoms} with those of the modified RWA EOMs in
Eq.~\eqref{eq:2ls_eoms_pulse}, showing the excited state population dynamics
for an isolated SQD ($R\rightarrow\infty$) interacting with a picosecond
and femtosecond pulse of area
$5\pi$ (according to Eq.~\eqref{eq:theta_sech}).
Fig.~\ref{fig:rwa_compare}~(a) shows $\rho_{22}(t)$ for a 1000-cycle pulse
($\tau_p\approx 0.11$~ps) and we can see that the RWA in this case provides an
adequate description of the dynamics, with population inversion occurring at
the end of the pulse as expected for a $5\pi$ pulse. The inset shows a
magnified region in
which we can see the effect of the RWA neglecting the quickly oscillating
terms: however, in the picosecond time-scale, these effects have negligible
influence on the overall dynamics.
Fig.~\ref{fig:rwa_compare}~(b) shows $\rho_{22}(t)$ for a 10-cycle pulse
($\tau_p\approx 1.1$~fs) where the pulse duration is of a comparable time-scale
to $2\pi/\omega_L$. In this case, we can see that the quickly oscillating terms
neglected in the RWA solution have a more significant effect on the overall
dynamics: importantly, complete population inversion is not achieved at the end
of the pulse, and there is a much more oscillatory behaviour.

In Ref.~\onlinecite{yang_ultrafast_2015}, a numerical solution to
Eq.~\eqref{eq:eoms} for pulsed excitation in SQD-MNP systems beyond the RWA is proposed. In
deriving Eq.~\eqref{eq:2ls_eoms_pulse}, $\esqd(t)$ is expressed by separating
out the positive and negative frequency parts as
\begin{equation}
    \esqd(t) \approx \frac{\hbar}{\mut}\left[ \left( \frac{\Omega(t)}{2}e^{-\imagn\omega_L t} +
    G\rho_{21}(t) \right) + \text{c.c.} \right] .
    \label{eq:esqd_t}
\end{equation}
Instead of invoking the usual RWA to arrive at
Eq.~\eqref{eq:2ls_eoms_pulse}, Yang et al. numerically solve
Eq.~\eqref{eq:eoms} using as $\esqd(t)$ the field in Eq.~\eqref{eq:esqd_t} (we
shall call this method the effective field method). However, in
deriving Eq.~\eqref{eq:esqd_t}, one must first separate out the slowly oscillating
components of the off-diagonal density matrix elements as in
Eq.~\eqref{eq:rho_bar} (see e.g. Ref.~\cite{artuso_strongly_2010}) and the
slowly-varying envelope approximation must also be used. Thus, while the
quickly oscillating terms are included, improving over the RWA, the pulse
duration must still be longer than $2\pi/\omega_L$. We shall presently
demonstrate how these assumptions render this approach unreliable for few-cycle
pulses when the interparticle distances are small.

\onecolumngrid

\begin{figure}[ht] 
    \begin{minipage}{\columnwidth}
            \includegraphics[width=0.45\columnwidth]{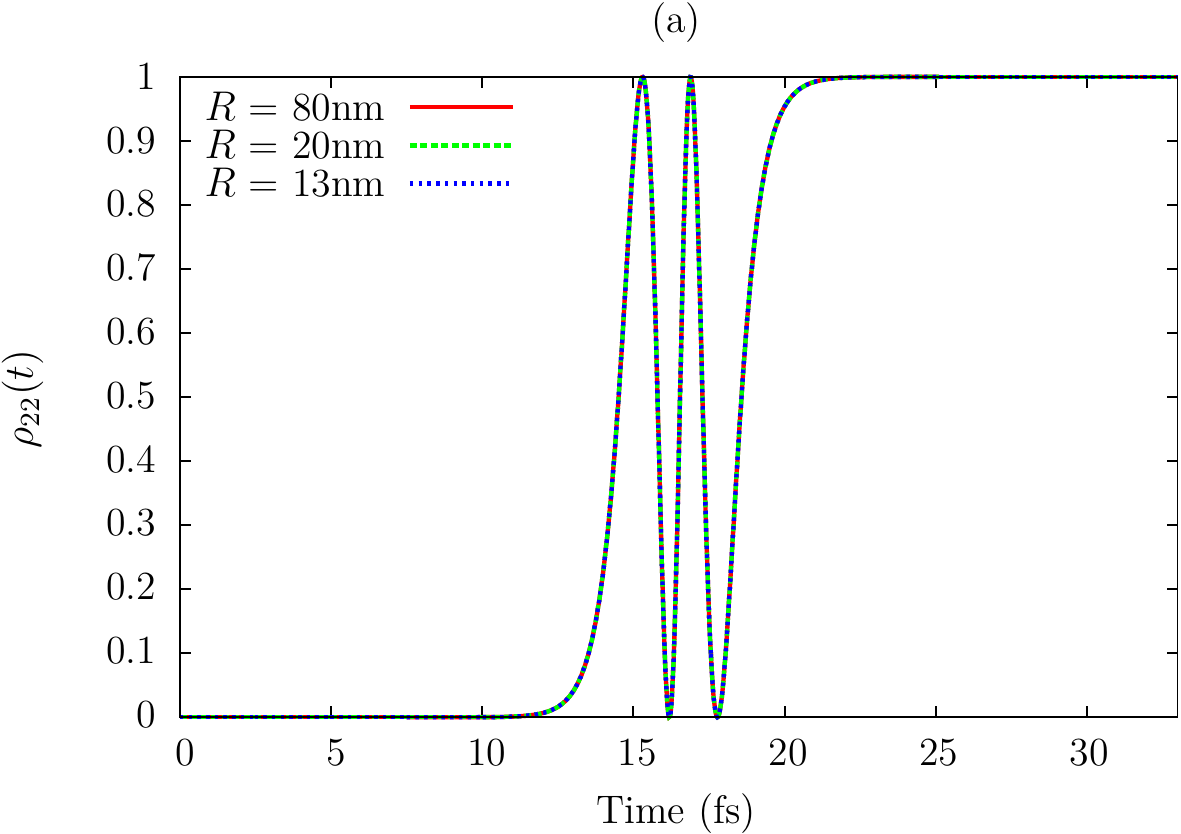}
            \hspace{0.25cm}
            \includegraphics[width=0.45\columnwidth]{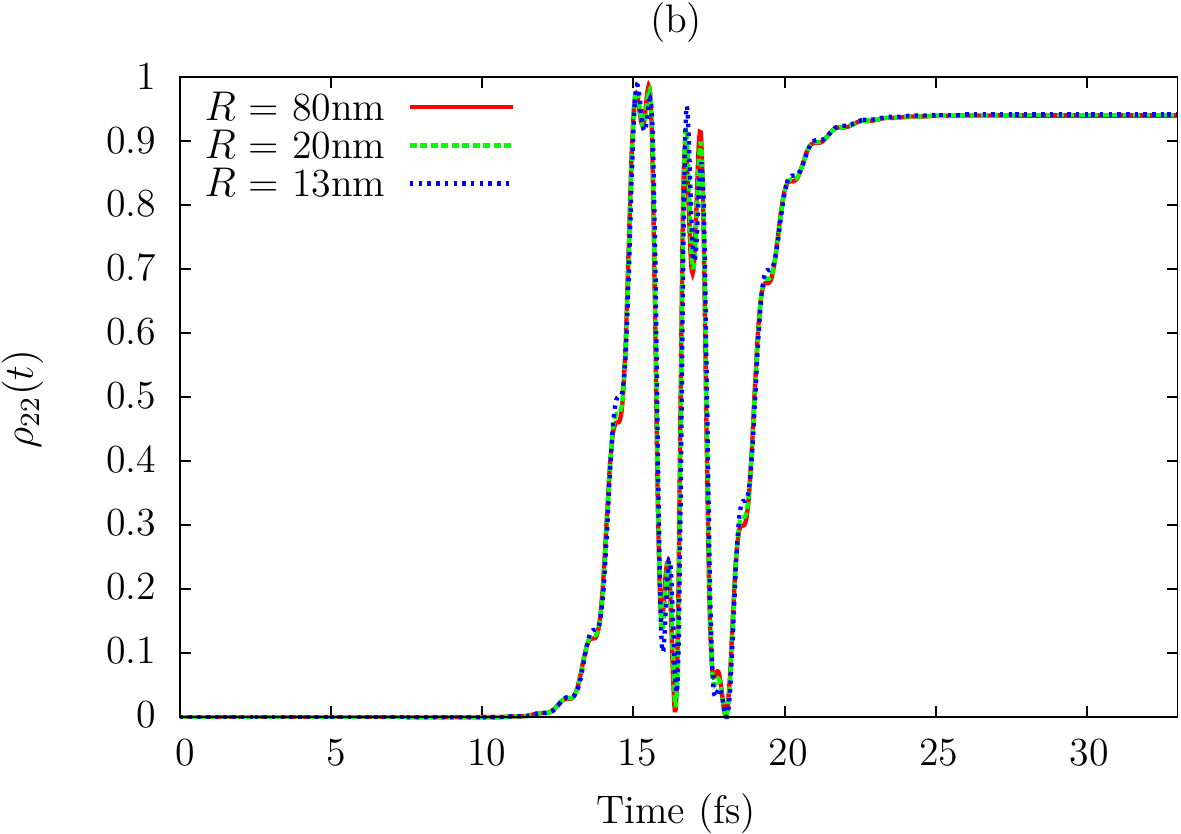}
            \vspace{0.25cm}
    \end{minipage}
    \begin{minipage}{\columnwidth}
            \includegraphics[width=0.45\columnwidth]{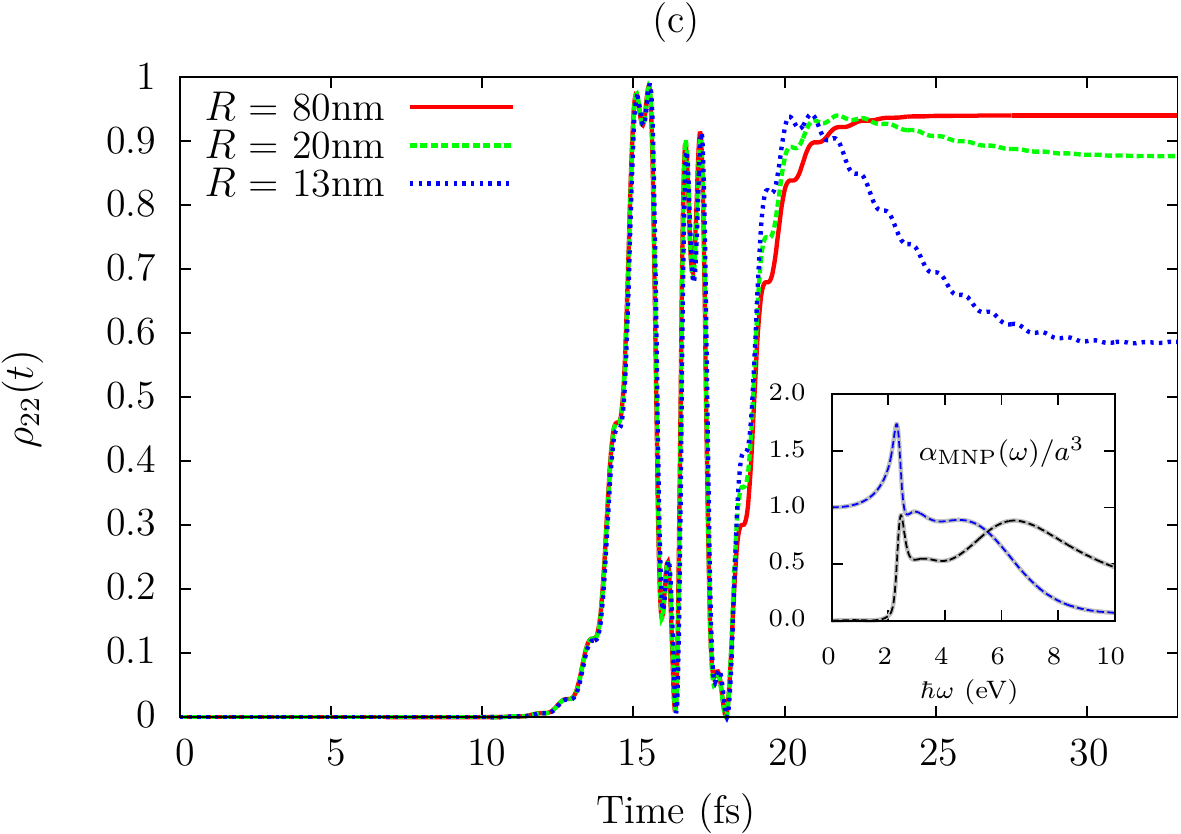}
            \hspace{0.25cm}
            \includegraphics[width=0.45\columnwidth]{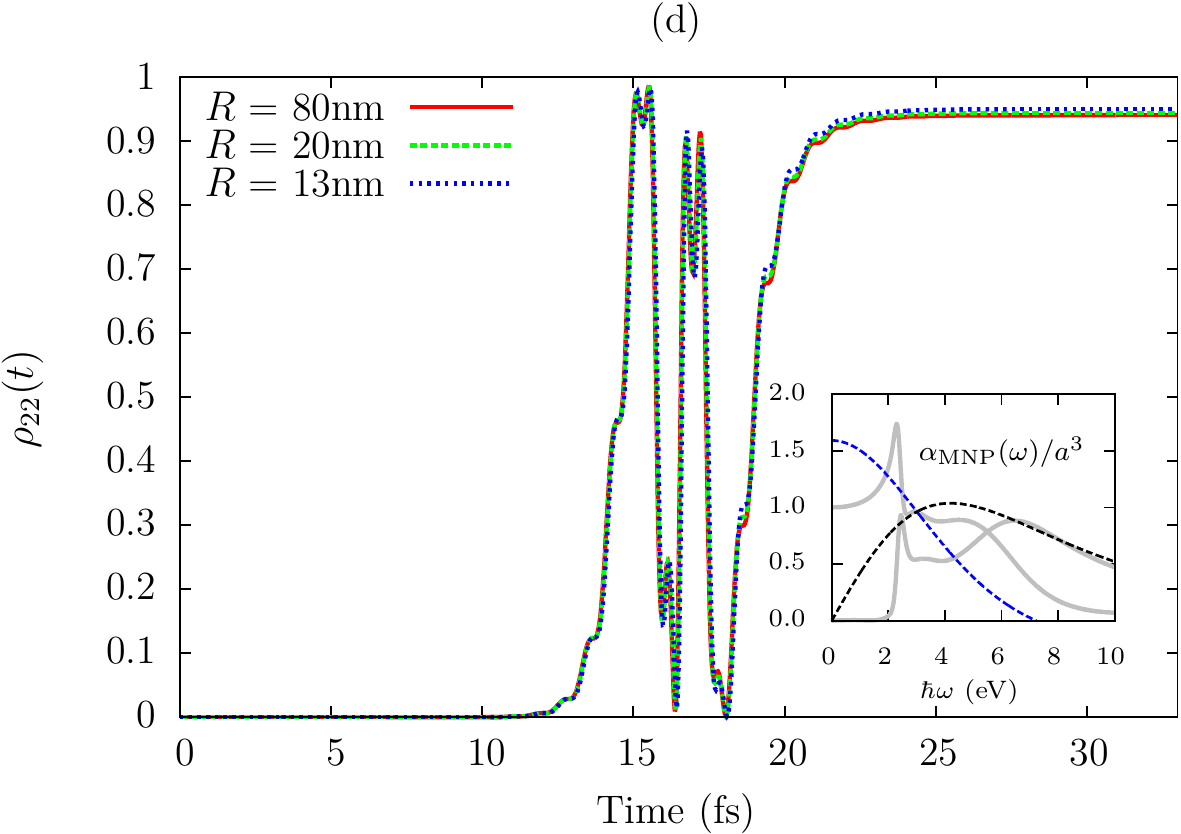}
    \end{minipage}
    \caption{Excited state population dynamics for an SQD-MNP system interacting
    with a 10-cycle sech pulse for various interparticle distances, $R$. The
field amplitude is chosen to give a $5\pi$ pulse area for each value
of $R$ according to Eq.~\protect\eqref{eq:theta_sech_R}. 
(a) Solution to the modified EOMS in Eq.~\protect\eqref{eq:2ls_eoms_pulse} under the
RWA and assuming a slowly-varying pulse envelope. 
(b) Solution to the original EOMs in Eq.~\protect\eqref{eq:eoms} beyond the RWA using
the effective field in Eq.~\protect\eqref{eq:esqd_t} which assumes a slowly-varying
pulse envelope. 
(c) Solution to Eq.~\protect\eqref{eq:eoms} using the PEOM method where the auxiliary
parameters are obtained by fitting $\alphamnp(\omega)$ in
Eq.~\protect\eqref{eq:chi_sphere} accurately over the range 0--10~eV using $N=21$
fitting functions.
(d) Same as (c) but where $\alphamnp(\omega)$ is fitted over a small range
close to $\hbar\omega_0$ (2.455--2.545~eV) using $N=1$ fitting functions.
Insets: real (blue dashed) and imaginary (black dashed) parts of the fitted
polarizability, $\alphamnp(\omega)/a^3$ (exact function shown in
grey).}
\label{fig:rwa_peom_compare}
\end{figure}    

\twocolumngrid

In Fig.~\ref{fig:rwa_peom_compare}, we compare the solutions for $\rho_{22}(t)$
based on the RWA, effective field method and the PEOM method. In each
case, the SQD-MNP system interacts with a 10-cycle sech pulse of area $5\pi$
with $R$-dependent amplitude given by Eq.~\eqref{eq:theta_sech_R} and the
excited state population dynamics are shown for various interparticle
distances. 

In Fig.~\ref{fig:rwa_peom_compare}~(a), the modified
RWA EOMS in Eq.~\eqref{eq:2ls_eoms} are solved and we can see that complete population
inversion occurs at the end of the pulse and the dynamics are identical for
each $R$ as expected from Ref.~\onlinecite{paspalakis_control_2013} and
Eq.~\eqref{eq:theta_sech_R}. 

In
Fig.~\ref{fig:rwa_peom_compare}~(b), the original EOMS in Eq.~\eqref{eq:eoms}
are solved beyond the RWA by taking $\esqd(t)$ of the form in
Eq.~\eqref{eq:esqd_t} similar to the calculations performed in
Ref.~\onlinecite{yang_ultrafast_2015}. In this case we see that the dynamics
are almost identical to the isolated SQD as shown in
Fig.~\ref{fig:rwa_compare}~(b) where the original EOMs are solved exactly.
At difference with the RWA solution in Fig.~\ref{fig:rwa_peom_compare}~(a),
complete population inversion does not occur as a consequence of the
RWA-breakdown.
On the other hand, the dynamics remain independent of $R$ as predicted by
Ref.~\onlinecite{paspalakis_control_2013}. We note at this point that
Ref.~\onlinecite{paspalakis_control_2013,yang_ultrafast_2015} employ a multipole description for the MNP response while our calculations use the simpler
dipole model. However, we have compared results using the same multipole
approximation and noticed no difference due to the short time-scales involved
here.

\begin{figure}[ht]
    \centering
    \includegraphics[width=\columnwidth]{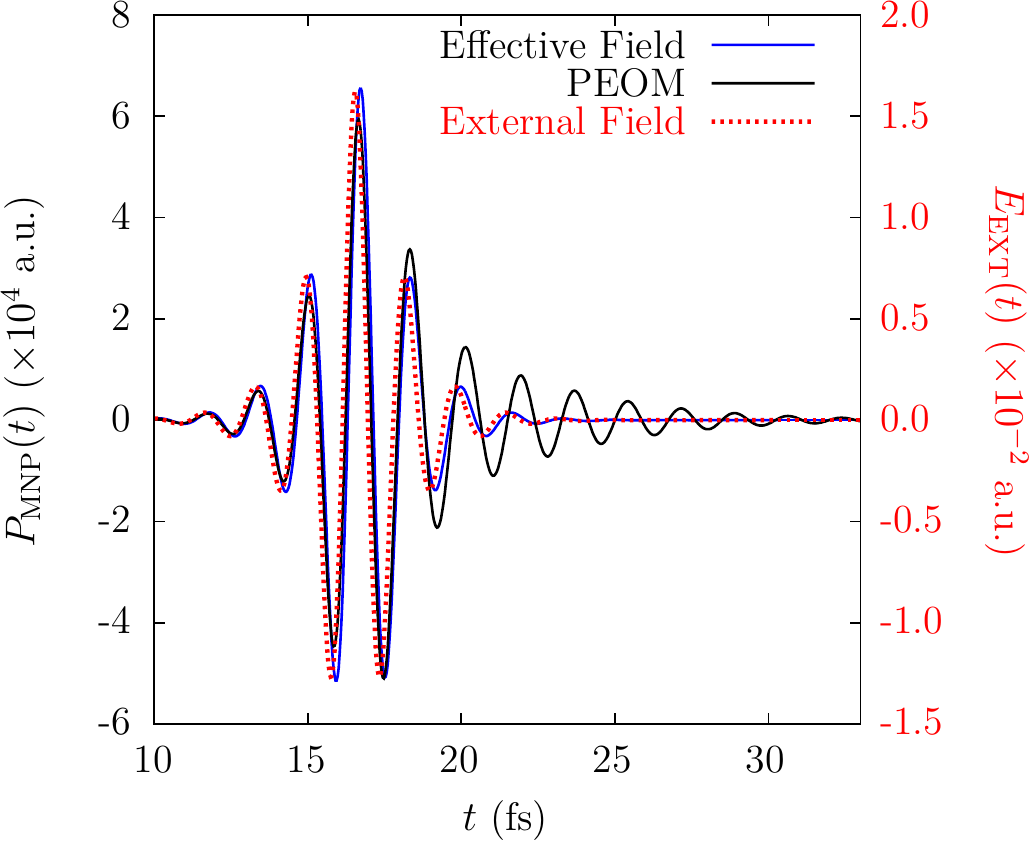}
    \caption{
        Time evolution of the MNP dipole moment, $\pmnp(t)$, for the
        $R=13$~nm cases in Fig.~\ref{fig:rwa_peom_compare}~(b) (solid blue) and (c)
    (solid black) where the effective field method and PEOM method are used
respectively. The corresponding external field is shown in dashed red.}
    \label{fig:pmnpt}
\end{figure}

In Fig.~\ref{fig:rwa_peom_compare}~(c), we solve the original EOMs in
Eq.~\eqref{eq:eoms} using the PEOM method. We obtain the auxiliary parameters
describing the MNP dipole moment by fitting $\alphamnp(\omega)$ in
Eq.~\eqref{eq:chi_sphere} to the functions in Eq.~\eqref{eq:chi_approx}. This
is achieved by a least-squares fit over the range 0--10~eV using $N=21$ fitting
functions to gain a fit of sufficient accuracy (see inset).
We see that for large $R$ ($R=80$~nm),
$\rho_{22}(t)$ resembles the results in (b).
However, as the interparticle distance decreases, the dynamics change
considerably, with larger effect towards the end of the pulse. For each $R$, the dynamics are
similar up to around 18~fs by which point the pulse is almost over (see
Fig.~\ref{fig:pmnpt}~(b)). After this point, the population for each $R$
reaches the same maximum value (around 0.95), but at different times: $\sim23$~fs for $R=80$~nm,
$\sim21$~fs for $R=20$~nm and $\sim20$~fs for $R=13$~nm. The population then
decreases more steeply as $R$ decreases, reaching as low as 0.6 for $R=13$~nm.

We ascribe the different results obtained with the PEOM in
Fig.~\ref{fig:rwa_peom_compare}~(c) and the effective field method in
Fig.~\ref{fig:rwa_peom_compare}~(b) to the fact that femtosecond pulses ($\sim
10$ cycles) excite
a broad range of frequencies: in particular, much broader than the
sub-picosecond pulses ($\sim 100$ cycles) for which the effective field
method~\cite{yang_ultrafast_2015} was originally developed.
As stated earlier, in writing
Eq.~\eqref{eq:esqd_t}, $f(t)$ must be slowly varying and the off-diagonal
density matrix elements must also first be separated into slowly and quickly
oscillating components. These approximations indirectly force the MNP to
respond only as $\alphamnp(\omega_L)$ (i.e. only at the driving frequency), as
apparent in the definitions of $\omegaeff$ and $G$ in Eq.~\eqref{eq:omegaeff}.
However, the femtosecond pulse has a large bandwidth ($>1$~eV), thus exciting a
broad range of frequencies in the MNP response.  Moreover, $\alphamnp(\omega)$
changes significantly over this range close to $\hbar\omega_0=2.5$~eV due to the
formation of the plasmon peak and thus one would expect the resulting
time-dependent dipole moment, $\pmnp(t)$, (and therefore $\esqd(t)$) to be
modified compared with that for long pulses of smaller bandwidths.  In
Fig.~\ref{fig:pmnpt}, $\pmnp(t)$ is shown for the $R=13$~nm cases in
Fig.~\ref{fig:rwa_peom_compare} (b) and (c). We can see that in the effective
field method, the MNP responds in phase with the external field,
while in the PEOM method the dipole moment continues to propagate well after
the pulse is over, thus contributing to the decline in population of the SQD.

\begin{figure}[ht]
    \centering
    \includegraphics[width=\columnwidth]{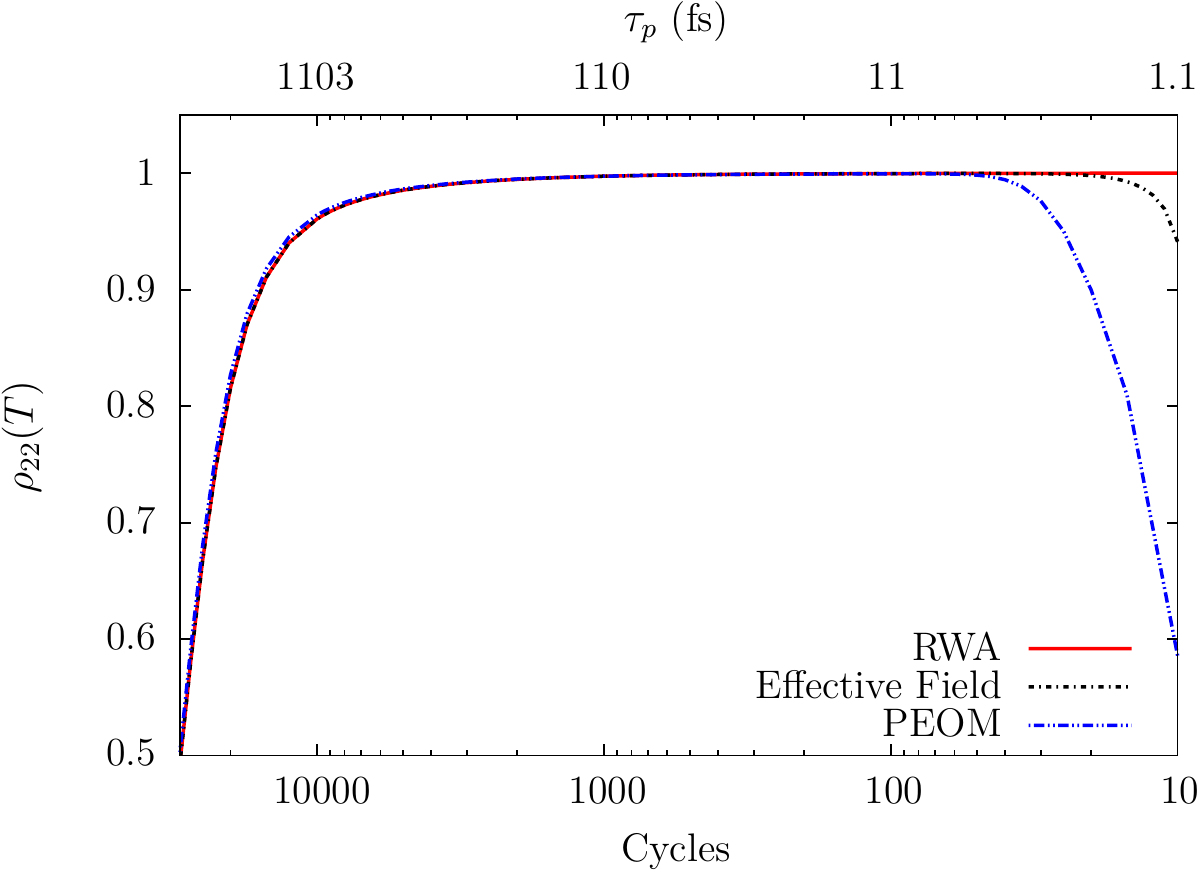}
    \caption{
        Value of $\rho_{22}(t)$ at the end of the pulse
for various pulse durations. The interparticle distance is $R=13$~nm and the
pulse area is chosen to be $5\pi$ according to Eq.~\protect\eqref{eq:theta_sech_R}.}
    \label{fig:change_length}
\end{figure}

We have stated that in the effective field method, the MNP responds only at the
driving frequency, $\omega_L$ (i.e. the polarizability is effectively constant,
$\alphamnp(\omega)\approx\alphamnp(\omega_L)$). This approximation is valid for
monochromatic waves (e.g. Eq.~\eqref{eq:plane_wave}) and for long pulses. On
the other hand, for very short pulses the frequency-dependence of the
polarizability is important due to the larger bandwidth. The PEOM method
overcomes this limitation as $\alphamnp(\omega)$ can be fitted over an
arbitrary frequency range (as in Fig.~\ref{fig:rwa_peom_compare}~(c)) so that
the relevant frequencies can be included in the dynamics. The constant
polarizability of the effective field method can be simulated within the PEOM
method by choosing a single, broad function ($N=1$) which agrees with
$\alphamnp(\omega)$ at $\omega_L$ and is approximately constant over the pulse
bandwidth region (see inset of Fig.~\ref{fig:rwa_peom_compare}~(d)).
Fig.~\ref{fig:rwa_peom_compare}~(d) then shows that the effective field results
from Fig.~\ref{fig:rwa_peom_compare}~(c) are indeed recovered.

Fig.~\ref{fig:change_length} summarizes the results and shows the range of
pulse durations for which the different approximations are valid. The three
different methods agree for $n>100$ cycles ($\tau_p>10$~fs). The effective field method
correctly describes the fall in final population as the pulse duration approaches
10 cycles due to the breakdown of the RWA, but we can see that when the full
response of the MNP is taken into consideration in the PEOM method, the effect
is much more enhanced. 

Overall, the results of this section show that when examing the response of
ultrashort pulses (fewer than $\sim 60$ cycles), one cannot rely on the RWA or
on the assumption that the MNP polarizability ($\alphamnp$) responds only at
the driving frequency, $\omega_L$. Therefore, one should consider more advanced
approaches. The PEOM methods is a valid alternative as it is not bound by such
approximations yet still its computational cost and complexity is similar to
that of, e.g., the RWA or effective field method.

\section{Conclusions\label{sec:conclusions}}
We have described a transferable hybrid approach to the electron dynamics of
a quantum system dipolarly coupled to a larger environment that can be treated
classically. This hybrid approach is based on a robust
projected equations of motion (PEOM) formalism.
The capabilities of the proposed hybrid approach have been demonstrated for the widely studied
case of a semiconductor quantum dot (SQD) coupled to a metallic nanoparticle (MNP).
The SQD has been modelled as a two-level system, while a semi-empirical model of the 
MNP susceptibility has been used.
We have validated this hybrid approach against both analytical and
semi-analytical benchmarks of the SQD-MNP response to
picosecond laser pulses, i.e., longer than $2\pi/\omega_0$, where
$\hbar \omega_0$ is the SQD energy gap. 
This is the regime of validity of the rotating wave approximation (RWA). 
However, the validity of the PEOM does not rely on either the RWA 
or improvements on it (e.g., the effective field method\cite{yang_ultrafast_2015})
and we have also modeled the response to femtosecond laser pulses.
In this regime, we have shown that the response of the SQD-MNP
is strongly affected by the details of the MNP susceptibility.
By artificially ``blurring'' the details of the MNP susceptibility,
the results of the hybrid approach
match the prediction of the effective field method.
To this extent, the proposed hybrid approach is inherently more accurate than
the other methods which rely on the RWA and improvements on it.

Beyond the validation for a two-level system, the  
PEOM formalism can be used for systems 
with an arbitrary number of levels
and is independent from the theoretical framework used to model the quantum system, e.g., the SQD.
In this work, we have used a density matrix approach, but the PEOM can be easily formulated
within a time-dependent density-functional theory framework, 
or the recently devised real-time approach to the Bethe-Salpeter
equation~\cite{attaccalite_real-time_2011} formalism.

The proposed hybrid approach shares similarities with other hybrid methods 
\cite{coomar_near-field:_2011,gao_communication:_2013,sakko_dynamical_2014}
and, in principle, can be also coupled to a finite-difference time-domain (FDTD)
description of the electromagnetic field.
On the other hand, an accurate FDTD model is less crucial if the SQD and MNP are 
sufficiently far apart. Moreover, the simpler dipolar coupling used in this work
is still popular~\cite{Li:16,terzis_2016,yang_ultrafast_2015} and
the PEOM formalism provides a necessary improvement as attention turns towards ultrafast
phenomena.

When both the near-field response and the electromagnetic 
scattering can be safely neglected, 
the proposed hybrid method provides a
computationally less expensive alternative to those more accurate approaches
which include an FDTD model of the electromagnetic field. 
This hybrid approach is also easier to integrate
into existing electronic structure codes, including
codes which employ periodic-boundary conditions. This is 
particularly relevant for modelling extended 
quantum systems (e.g., two-dimensional
semiconductors) coupled to MNPs.~\cite{eda_two-dimensional_2013}
%

%
%
\begin{acknowledgments}
    RM acknowledges financial support from the UK Engineering and Physical
    Sciences Research Council.
\end{acknowledgments}

\bibliography{Master}

\end{document}